\documentclass{elsarticle}
\usepackage{epsfig}

\begin{document}
\begin{frontmatter}
\title{
Potential fitting biases resulting from grouping data into variable width bins
}

\author[st]{S.Towers}
\ead{smtowers@asu.edu}
\address[st]{Arizona State University, Tempe, AZ, USA}

\begin{abstract}

When reading peer-reviewed scientific literature describing any analysis
of empirical data, it is natural and correct 
to proceed with the underlying assumption that experiments have made good faith efforts to
ensure that their analyses yield unbiased results.
However, 
particle physics experiments are expensive and time consuming to carry out,  
thus if an analysis has inherent bias (even if unintentional),
much money and effort can
be wasted trying to replicate or understand the results,
particularly if the analysis is fundamental to our understanding of the universe.

In this note we discuss the significant
biases that can result from data binning schemes.
As we will show, if data are binned such that
they provide the best comparison to a particular (but incorrect) model, the 
resulting model parameter estimates when fitting to the binned data can be 
significantly biased, leading us to too often
accept the model hypothesis when it is not in fact true.  
When using binned likelihood or least squares methods
there is of course no {\it a priori} requirement that data bin sizes need to be constant,  
but we show that fitting to data
grouped into variable width bins is particularly prone to produce biased results if the
bin boundaries are chosen to optimize the comparison of the binned data to a wrong model.
The degree of bias that can be achieved simply with variable binning can be surprisingly large.

Fitting the data with an unbinned likelihood method, when possible to do so,
is the best way for 
researchers to show their analyses are not biased by binning effects.
Failing that, equal bin widths should be employed as a cross-check of the
fitting analysis whenever possible.

\end{abstract}
\end{frontmatter}


\section{Introduction}

In 1830, Charles Babbage, Professor of Mathematics at Cambride University, wrote an 
now-famous
essay describing
the ways that scientific data can be 
consciously or unconsciously manipulated to achieve a desired conclusion\cite{bib:babbage}.
In it, he describes one of the manipulation methods 
as {\it ``retaining only those results that fit the theory, and discarding others''}.
Many would likely assume that this kind of manipulation only involves 
exclusion of data,  but in this paper we discuss a perhaps much more subtle example; 
excluding none of the data, but
rejecting data analysis methods that do not result in the desired conclusion.

As an example of how this can occur in practice, 
let us consider a hypothetical empirical data set that we wish to compare
to a model. The data are stochastic (ie; have some random variation), and thus
vary somewhat about the true mean. Let us refer to data that stochastically fluctuate 
higher than the true 
mean as $Y^{\rm high}$, and data that stochastically fluctuate lower as $Y^{\rm low}$.
Fairly often, just due to stochasticity, several events in succession may fluctuate high (or low).
Thus, for example, if we wish the data in a particular bin
to appear to be consistent
with a predicted mean hypothesis that is larger than the true mean, we can simply adjust
the bin width and optimize the bin boundaries to increase the ratio of $Y^{\rm high}$ to $Y^{\rm low}$ 
events in the bin.  

If we are attempting to compare a longer time series of data to the predictions
of a desired (but incorrect) model, the optimization process
becomes somewhat more complicated because several bin boundaries must be simultaneously
optimized to take advantage
of stochastic fluctuations such that as many bins as possible agree with the incorrect
model hypothesis.  However, straightforward Monte Carlo methods can be employed to achieve this, where many
different random combinations of bin boundaries are tested, choosing the boundaries that achieve the 
best fit of the binned data to the incorrect model. 

\begin{figure}[h]
   \begin{center}
    \begin{minipage}[t]{1.0\linewidth}

      \vspace*{-3cm}  
      \hspace*{-3cm}
      \epsfig{file=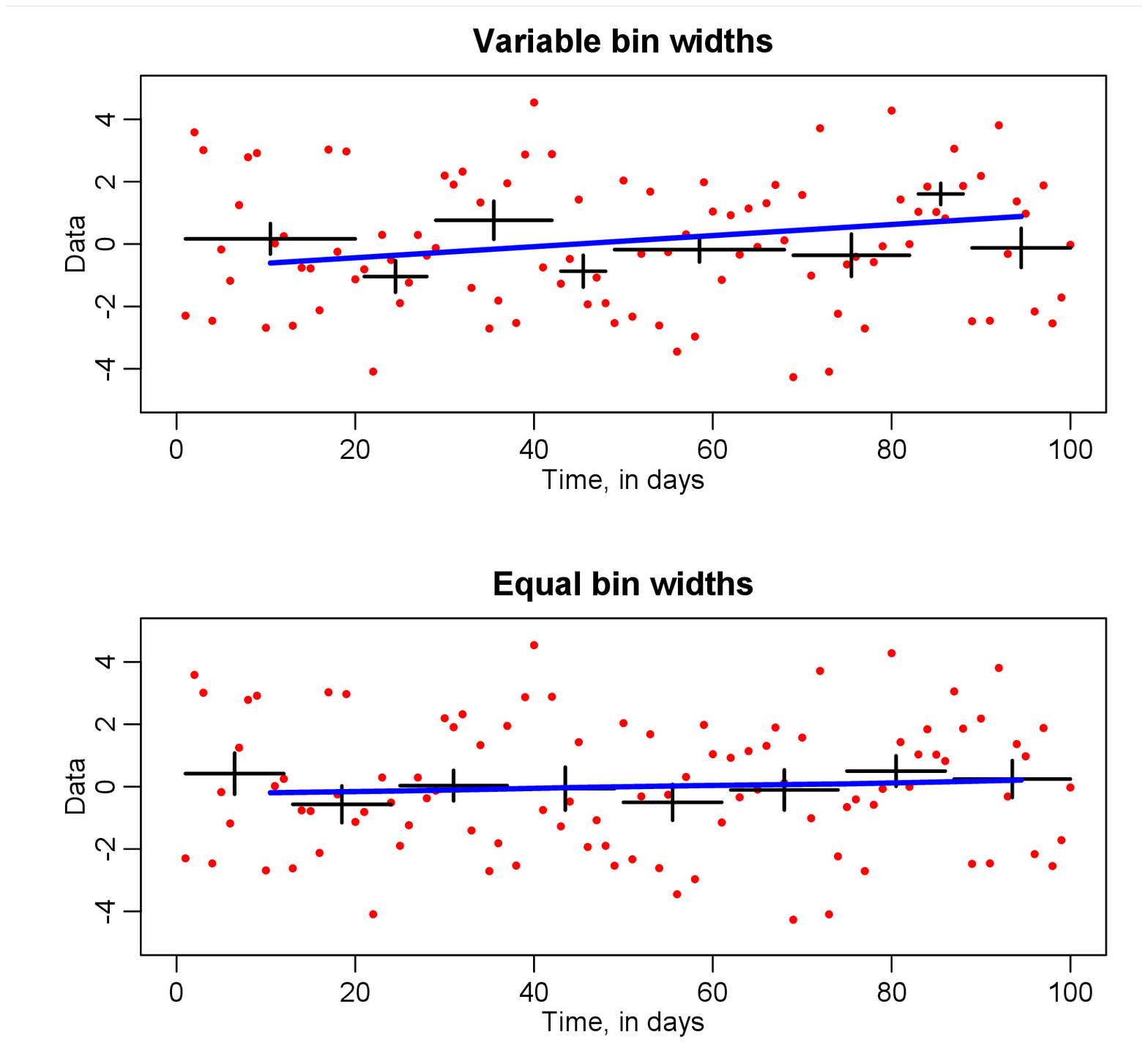, width=18cm}
    \end{minipage}
    \vspace*{-7cm}
  \caption{
      \label{fig:trend}
As an example of the significant biases that can result when fitting with variable width bins, 
data are generated over 100 days, sampling from a Normal distribution, with no
trend (red points).  The top plot shows the fit when variable width bins
are employed to optimize the agreement with the grouped data (black points) to
a model that assumes the slope is $0.02$ days$^{-1}$.  The fit returns an estimated
slope of $0.018\pm0.006$, which is three standard deviations from the true value.
The bottom plot shows the fit to the data grouped into bins of equal width.  The
fit yields an estimated slope of $0.005\pm0.007$, consistent with the true value.
   }
   \end{center}
 \end{figure}

In Figure~\ref{fig:trend} we give a simple example of this kind of manipulation.  A time
series of data recorded over 100 days
is Monte Carlo generated with a Normal distribution of zero mean and no trend (red points).  
However, we (wrongly) believe that the underlying model has trend with slope $0.02$ days$^{-1}$.  
Using varying bin widths from 5 to 20 days,
we thus use Monte Carlo methods to determine the bin boundaries that yield 
a binned likelihood fitted trend that is most consistent with this mistaken hypothesis.  The
results are shown in Figure~\ref{fig:trend}; the fitted slope with the optimized binning
is $0.018\pm0.006$, 
which is nicely consistent with the desired slope of $0.02$,
and is three standard deviations away from the true slope of zero.
However, if equal sized bins are employed, the likelihood fit
returns a slope of $0.005\pm0.007$, which is consistent with zero, and
over two standard deviations away from the
desired incorrect slope. 

Thus, when given the freedom to chose variable bin widths and the placement of the bin boundaries,
we see from just this one simple example that it is possible for researchers 
to accept a desired (but wrong) hypothesis, and reject the true hypothesis.
In the following sections we will discuss a much more complicated example that 
involves fitting the phase of periodic data with carefully chosen variable bin widths.
We will show that if empirical data exhibit annual modulation due to
a true underlying phase, $\phi_a$, implementing variable width binning when fitting
for the phase can
unfortunately often result in an incorrect conclusion that the
the data are instead consistent with a different phase, $\phi_b$.

Before moving on, we would like to stress  
that the manipulation of the data in this way, when it occasionally occurs, 
should not be assumed to be motivated
by deliberate researcher 
prevarication; rather, such manipulations can happen when the researchers sincerely believe 
an incorrect model to be the true underlying description
of their data, and thus wish to present the data in a way that highlights what they believe to be
evidence of that relationship.

\section{Methods}
\label{sec:methods}

Here we consider an illustrative example with basis
in the real world, simulating periodic data
similar to that of a typical dark matter experiment\cite{bib:hasenbalg}.
Daily data with annual periodicity
are simulated over a five year time span,
with an annual variation of amplitude $A\sim0.015$, and true phase $\phi_a$, which for
{\it illustrative purposes only} we
assume to be due to background or some other systematic effect.   
To simulate the observed data, $Y^{\rm obs}_i$, at day $t_i$, we thus use the model
\begin{eqnarray}
Y^{\rm obs}_i = A \cos{2\pi(t_i-\phi_A)} + \sigma_a {\cal N}(0,1),
\end{eqnarray} 
where ${\cal N}(0,1)$ is random number drawn from the Normal distribution.
For this example we assume the daily variation in the
data is $\sigma_a\sim0.05$.

We examine various binnings of the data, with bins between 30 days to 90 days in 
width.  
One thousand
randomly selected binning schemes are examined, and the binning scheme is
chosen that provides the best agreement of the fitted phase 
to the model 
\begin{eqnarray}
Y^{\rm pred}_i = A \cos{2\pi(t_i-\phi_b)} \hspace*{2cm} \phi_b\ne\phi_a,
\end{eqnarray}
and worst agreement to the model with the true phase, $\phi_a$.

Using this unequal width
binning scheme, we then perform a binned maximum likelihood fit to the data to obtain the estimated
phase, $\hat{\phi}$, and its uncertainty, $\Delta\hat{\phi}$.

We compare this to the phase estimate of an unbinned likelihood fit,
and we also examine the phase estimate of a binned likelihood fit with
bins of equal width.

Here we arbitrarily assume that the phase of physics
interest, $\phi_b$, is 90 days past January 1$^{\rm st}$.  We examine values of
$\phi_a$ from 45 to 85 days past January 1$^{\rm st}$, in steps of 5 days.
Note that it doesn't matter for this analysis what we assume for $\phi_b$, because the results
only depend on the absolute difference between $\phi_b$ and $\phi_a$.

For each true phase value,
$\phi_a$, we repeat the simulation and fitting procedure 250 times.

\section{Results}

\begin{figure}[h]
   \begin{center}
    \begin{minipage}[t]{1.0\linewidth}
      \vspace*{-3cm}  
      \hspace*{-3cm}
      \epsfig{file=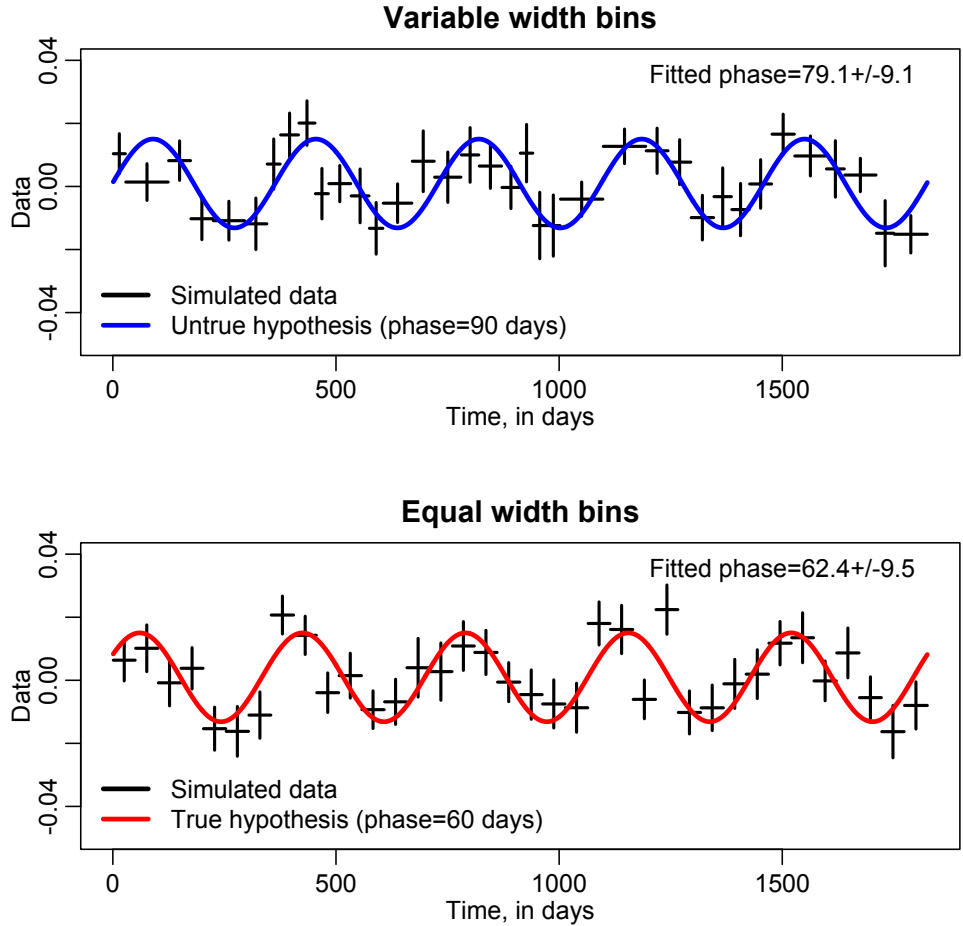, width=18cm}
    \end{minipage}
  \vspace*{-7cm}
  \caption{
      \label{fig:example}
As another example of the significant biases that can result when fitting with variable width bins, 
we fit to simulated data with an annual harmonic variation with true phase $\phi_a=60$ days (relative
to January 1$^{\rm st}$).
In the upper plot, variable binning is used to obtain the best comparison to an incorrect
model with
phase $\phi_b=90$ days.  The fit to the variably binned data yields an estimated phase consistent with this
untrue hypothesis, and statistically inconsistent with the true hypothesis.
Conversely, fitting with equal sized bins (bottom plot) yields an estimate of the phase that
is statistically consistent with the true hypothesis, and statistically inconsistent with the untrue
hypothesis.
The Pearson $\chi^2$ statistics for the upper and lower plots are $29$ and $40$,
 with 34 degrees of freedom, respectively.
   }
   \end{center}
 \end{figure}

An example of the significant bias that can be achieved with variable width bins when
fitting to the simulated harmonic data is shown in
Figure~\ref{fig:example}.  The true phase of the data is $\phi_a=60$ days and the variable width
binning is
tuned to best match a desired phase of $\phi_b=90$ days. 
The fit to the variably binned data yields an estimated phase,
$\hat{\phi}=79.1\pm9.1$,
which is consistent with the
untrue hypothesis,
and statistically inconsistent with the true hypothesis.
Conversely, fitting with the same number of
equal sized bins (bottom plot in Figure~\ref{fig:example}) 
yields an estimate of the phase,
$\hat{\phi}=62.4\pm9.5$, 
which
is statistically consistent with the true hypothesis 
and statistically inconsistent with the untrue
hypothesis.

\begin{figure}[h]
   \begin{center}
    \begin{minipage}[t]{1.0\linewidth}
      \vspace*{-3cm}  
      \hspace*{-3cm}
      \epsfig{file=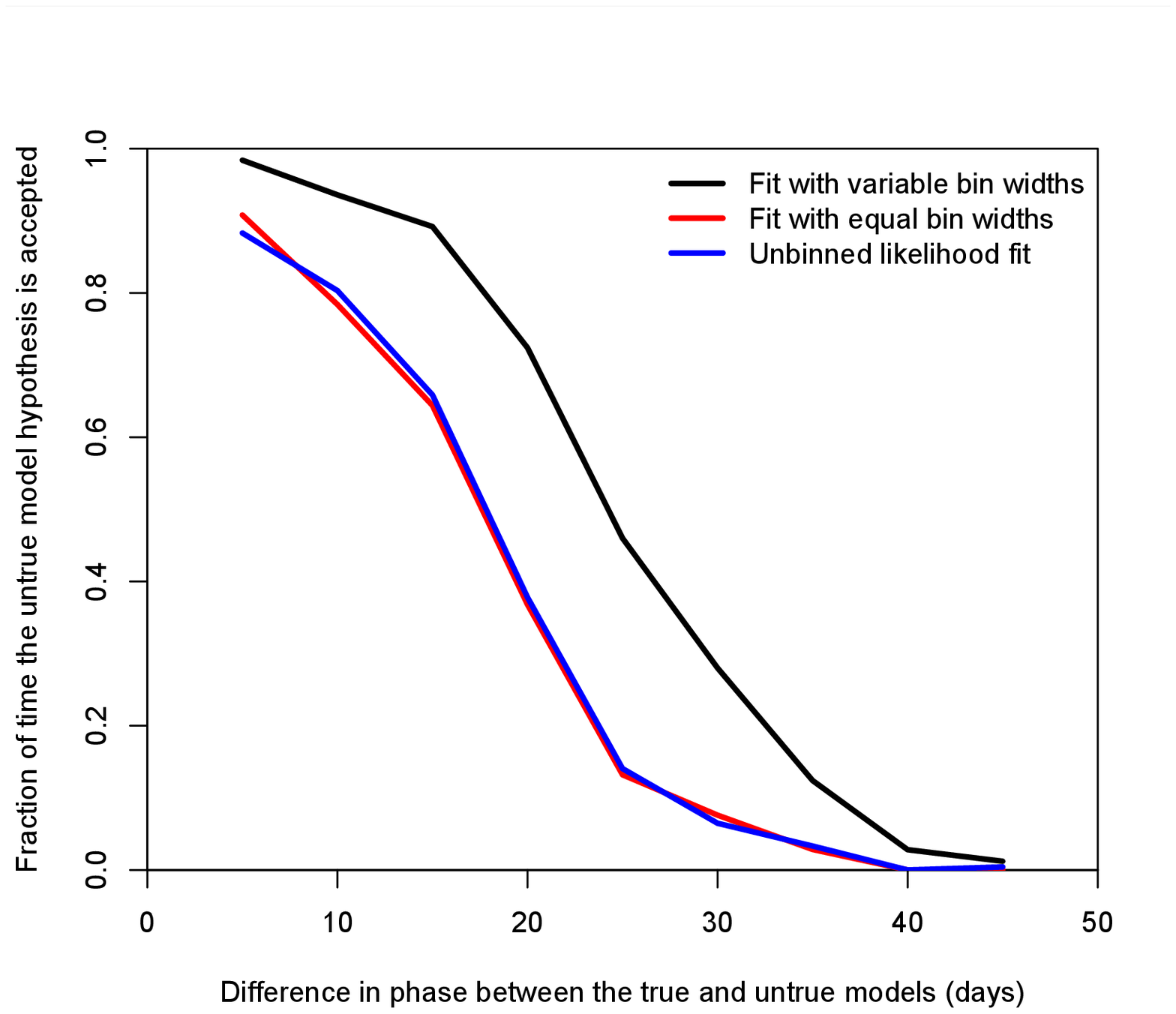, width=18cm}
    \end{minipage}
     \vspace*{-7cm}
  \caption{
      \label{fig:results}
For simulated data with annual periodicity, 
we examine 250 trials for true phase values, $\phi_a$, that
differ from a desired phase hypothesis, $\phi_b$, by various amounts between
5 to 45 days.
The blue line shows the results achieved
when an unbinned likelihood fit is employed.
The red line shows the results of a binned likelihood fit with bins of equal width.
Both types of fits yield values of $\phi_a$ that are unbiased.
The black line shows the results achieved with a binned likelihood fit with bins
of varying width, where the bin boundaries are chosen to optimize agreement with
the untrue phase hypothesis, $\phi_b$.
The fits with variable bin widths produce significantly biased results that cause
us to too often conclude the data are consistent with the untrue hypothesis.
   }
   \end{center}
\end{figure}

The results of using variable binning with 250 data simulations for various values of 
$\phi_a$ reveal that
on average 
the variably binned fit yields a 
phase estimate that is around one standard deviation
closer to the untrue hypothesis than the phase estimate from the unbinned and/or
equal bin width fit.
This results in the untrue phase hypothesis being accepted much more often than it
would be if the fit is unbiased.  For instance, as seen in Figure~\ref{fig:results},
when the true and untrue phase hypotheses are a month apart, using variable binning instead
of unbinned or equal width binning more than
triples the chances of accepting the untrue hypothesis as being correct (from 8\% to 28\%).
For a three week phase difference, the fit with variable binning doubles the probability
of accepting the untrue hypothesis from 37\% to 73\%.
Indeed, as seen in the Figure, 
for all differences between the phase hypotheses, variable binning yields a significantly 
improved probability of accepting the untrue hypothesis.

\section{Summary}

We have shown that fitting to data grouped into bins of variable widths can yield significantly
biased results.  Using a simple example of linear data, and a much more complicated example of 
periodic data, we have shown that fitting with variable binning can too often
produce biased estimates that
result in an incorrect model hypothesis being
accepted as true.   In the cases examined,
we have shown that unbinned likelihood fits and/or binned fits with
bins of equal width produce unbiased results.

The examples given in this paper are illustrative only, and certainly not
an exhaustive examination of the degree of bias that can be obtained in
any particular physics analysis when variable width binning is employed.
Researchers must thus be careful to employ variable bin widths only when absolutely
necessary, and reviewers must be careful to inquire about the underlying motivation when
reviewing a fitting analysis wherein the bin widths are variable.
The only truly reliable cross-check of sensitivity to binning effects is to fit with
equal width bins or, ideally, perform an unbinned likelihood fit.  It is not enough for
researchers to merely state that ``alternate'' binning schemes were examined,
since the alternate binning
schemes may have been chosen to be different, but still nevertheless quite optimal in
affirming the desired (but wrong) hypothesis.


\end{document}